\documentclass[a4paper,12pt]{JHEP3}
\usepackage{amssymb,amsmath}

\usepackage{epsfig}
\usepackage{graphics}
\usepackage{subfigure}

\usepackage{amsfonts}

\newcommand{\tr}{\mathop{\rm tr}\nolimits}

\newcommand{\ii}{{\rm i}}

\newcommand{\half}{{\scriptstyle\frac{1}{2}}}

\title{Chains of Skyrmions}
\author{Derek Harland\footnote{email address: d.g.harland@durham.ac.uk},
   R. S. Ward\footnote{email address: richard.ward@durham.ac.uk}
  \bigskip
  \\Deparment of Mathematical Sciences,
  \\Durham University,
  \\DH1 3LE}

\abstract{
Skyrme chains are topologically-nontrivial solutions of the Skyrme model
which are (quasi-)periodic in one spatial direction. We report numerical
and analytic investigations which show that such solutions exist. Chains
of 1-skyrmions are
reasonably well approximated both as parallel vortex-antivortex pairs,
and in terms of the holonomy of Yang-Mills calorons. As the period
increases, the 1-skyrmions clump together, for example giving chains of
2-skyrmions or 4-skyrmions.
}

\begin{document}

\maketitle



\newpage


\section{Introduction}

The Skyrme model is a three-dimensional field theory with topological
soliton solutions, which is believed to provide a good model of nuclear
physics; the topological solitons (skyrmions) are identified with baryons.
In this paper we investigate skyrmion chains,
in other words topological solutions which are
periodic (or quasi-periodic) in one of the spatial directions.

Isolated skyrmions have been well-studied, up to relatively high charge
\cite{BS02}. Multiply-periodic configurations have also been considered,
for example a ``Skyrme domain wall'' solution (doubly-periodic)
\cite{BS98}, and a ``Skyrme crystal'' (triply-periodic)
\cite{KS88, KS89, CJJVJ89}.  However, to date no one
has considered topologically-nontrivial chains. There have been studies
of cylindrically-symmetric skyrme strings \cite{J89, NS08}, but these are
topologically trivial.

The field equations of the Skyrme model are non-linear, and no exact analytic
solutions are known; so the only direct means of studying solutions is by
numerical simulation.  Since the system is three-dimensional, numerical methods
are not fast.  However, a number of alternative methods are known: these
typically involve a relatively simple approximation, which is empirically
observed to approximate the true (numerically-determined) behaviour of the
solutions. The three most prominent ans\"{a}tze of this type are the product
ansatz (see for example \cite{Sch94}), the Atiyah-Manton construction
\cite{AM89},
and the rational map ansatz \cite{HMS98}.  We have been able to adapt the first
two to study chains, but the rational map ansatz appears unsuited to this task.

The plan of the paper is as follows. Section 2 contains a discussion of
the topology of skyrmion chains; section 3 has results on adapting the
Atiyah-Manton construction to obtain approximate skyrmion chains from
Yang-Mills calorons; section 4 discusses the approximation of skyrmion chains
in terms of parallel vortex-antivortex pairs; and section 5 contains some
full 3-dimensional numerical results.

We show that there is a 1-skyrmion chain with a preferred period (the period
which minimizes the energy-per-period of the chain), and that the
caloron-generated field is a good approximation to this solution. The
vortex-antivortex pair is not as good, but becomes more accurate for denser
(lower-period) chains. Generally speaking, a chain of 1-skyrmions is
unstable for larger periods, in the sense that the individual 1-skyrmions
tend to clump into higher-charge skyrmions. We give some numerical
results which illustrate this, in particular showing the appearance of
chains of 2-skyrmions and of 4-skyrmions.


\section{The Skyrme model and chains}

The static Skyrme field $U$ is defined on $\mathbb{R}^{3}$, and takes
values in SU(2). Defining $L_i=U^{-1}\partial U/\partial x^i$, we take the
energy density of $U$ to be
\begin{equation}\label{skyrme energy density}
  \mathcal{E}:=-\frac{1}{2}\mbox{Tr}(L_i L_i)-
     \frac{1}{16}\mbox{Tr}([L_i,L_j][L_i,L_j]);
\end{equation}
and the normalized energy of $U$ on $\mathbb{R}^{3}$ is
\begin{equation}\label{skyrme energy}
  E:=\frac{1}{12\pi^2}\int_{\mathbb{R}^3} \mathcal{E} dx^1 dx^2 dx^3.
\end{equation}
The boundary condition at spatial infinity is $U\to1$ as $r\to\infty$,
where 1 denotes the identity element of SU(2). So topologically, a
Skyrme field defines a map $S^3\rightarrow SU(2)$, and such a
map has a degree $B\in\mathbb{Z}$.  This topological charge can be computed by
the integral
\begin{equation}
  \label{skyrme charge}
  B = \int_{\mathbb{R}^3} \mathcal{B}\, dx^1\, dx^2\, dx^3,
\end{equation}
where
\begin{equation} \label{skyrme charge density}
  \mathcal{B}= \frac{1}{24\pi^2} \epsilon_{ijk} \mbox{Tr}(L_i L_j L_k)
\end{equation}
is the topological charge density. The energy (\ref{skyrme energy}) satisfies
the topological lower bound $E\geq B$.
Finite-energy fields $U$ which locally minimize $E$ will be called skyrmions.
The symmetry group of the system (including the boundary condition) consists
of the spatial translations, the spatial rotations ${\rm O(3)_{sp}}$, and the
iso-rotations ${\rm O(3)_{iso}}$.

A Skyrme chain with period $\beta>0$ and relative orientation
$R\in {\rm SO(3)_{iso}}$ is a finite-energy Skyrme field
$U:\mathbb{R}^3\rightarrow SU(2)$ satisfying
\begin{equation} \label{orientation}
  U(x,y,z+\beta) = R\cdot U(x,y,z),
\end{equation}
minimizing the energy functional
\[
  E = \int_0^{\beta} \int_{\mathbb{R}^2} \mathcal{E} dx\,dy\,dz,
\]
and satisfying the boundary condition $U\to1$ as $x^2+y^2\to\infty$.
One may think of such a configuration as a chain of equally-spaced
skyrmions along the $z$-axis, with each skyrmion being iso-rotated by $R$
relative to its neighbours. In the Appendix, it is shown that the quantity
\[
   B = \int_0^{\beta} \int_{\mathbb{R}^2} \mathcal{B}\, dx\,dy\,dz
\]
is necessarily an integer, and this integer is called the
\emph{topological charge}
of the chain. (If $R$ is the identity, then it is already clear that $B$ is
an integer, since then the field $U$ is in effect a map from $S^2\times S^1$
to SU(2), and $B$ is the degree of this map.)

A single skyrmion has the spherically-symmetric hedgehog form
$U(\mathbf{x})=\exp\left\{ {\rm i} f(r) \hat{x}^j \sigma^j \right\}$,
where $\sigma^j$ denotes the Pauli matrices. Consequently, the
iso-rotations $R\in{\rm SO(3)_{iso}}$ of a single skyrmion may be identified
with the spatial rotations, and we may think of the relative orientation
of two skyrmions as being a relative orientation in physical space.
If we have two well-separated 1-skyrmions, then it is well-known \cite{Sch94}
that the force between them depends on this relative orientation:
\begin{itemize}
  \item If the two skyrmions have the same orientation, then they repel.
    This case is called \emph{aligned}.
  \item The strongest repulsive force occurs when one of the skyrmions is
   rotated by $\pi$ about the line $l$ joining them. This case
   is called the \emph{repulsive channel}.
  \item The strongest attractive force occurs when one of the skyrmions is
    rotated by $\pi$ about an axis perpendicular to $l$. This case is
    called the \emph{attractive channel}.
\end{itemize}
The symmetry of such a 2-skyrmion configuration can be inferred from the
dipole model of skyrmions \cite{MS04}, or equivalently by looking at a
superposition of hedgehog configurations.
A pair of skyrmions which are aligned, or in the repulsive channel,
has an axial symmetry O(2) (consisting of rotations and reflections
which fix $l$); but a pair in the attractive channel has only a discrete
symmetry group $D_2$ (generated by reflections in two perpendicular planes
whose intersection is $l$).

Guided by this, we define three types of unit-charge skyrmion chain
as follows. Writing $U({\bf x})=\exp\left\{w({\bf x})\right\}$,
where $w$ takes values in the Lie algebra $su(2)$, we may regard
the relative orientation operator $R$ defined in (\ref{orientation})
as acting on $w({\bf x})$ via the adjoint action. Suppose that along
the $z$-axis, $w$ has the form $w=g(z) v$, where $v$ is a fixed
element of the Lie algebra $su(2)$, and $g(z)$ is a real-valued function.
We then identify the following three special cases:
\begin{itemize}
  \item A chain with $R=1$ is called \emph{aligned}.
  \item A chain for which $R$ is a rotation by $\pi$ about $v\in su(2)$,
    is called \emph{maximally-repulsive}.
  \item A chain for which $R$ is a rotation by $\pi$ about an axis
    perpendicular to $v\in su(2)$ is called \emph{maximally-attractive}.
\end{itemize}
As before, the maximally-attractive chain has only a discrete symmetry, whereas
the other two have a continuous axial symmetry. We expect that only the
maximally-attractive chain will have an energy less than that of an isolated
skyrmion, and it is this type that we shall concentrate on in what follows.


\section{Skyrme chains from calorons}

The Atiyah-Manton ansatz is method of generating approximate skyrmion
configurations on $\mathbb{R}^3$ by evaluating the holonomy of 4-dimensional
Yang-Mills instantons. The construction is topologically natural, in the
sense that the holonomy of an $N$-instanton is an $N$-skyrmion; and for an
appropriate choice of the instanton scale, the resulting skyrmion field
is a surprisingly good approximation to the actual skyrmion.

The method involves evaluating the holonomy of the gauge field along a family
of parallel lines in the 4-dimensional space. For example, we could choose the
family of lines parallel to the $x^0$-axis in $\mathbb{R}^4$, and then the
Skyrme field is obtained via the path-ordered exponential integral
\[
  U(x^1,x^2,x^3) = {\cal P} \exp \left( \int_{-\infty}^{\infty} A_0(x) dx^0 \right).
\]
In one periodic version of this construction which has been studied previously,
it was shown that one may obtain a good approximation to the Skyrme crystal
from instantons on the 4-torus \cite{MS95}.

If we want to obtain a Skyrme chain, then we should start with a gauge field
on $S^1\times\mathbb{R}^3$ satisfying
\begin{equation}\label{algebraic gauge}
  A_\mu (x^0+\beta, x^1,x^2,x^3) = R\cdot A_\mu(x^0,x^1,x^2,x^3),
\end{equation}
and integrate along a family of parallel lines perpendicular to the $x^0$-axis.
Calorons provide examples of gauge fields satisfying (\ref{algebraic gauge}).
Let $A_\nu$ be the gauge potential of a caloron, strictly-periodic with period
$\beta$, with instanton charge equal to 1,
and with vanishing monopole charge (see \cite{GPY81, N03} for details).
So near infinity, $A_0$ does not wind, and has the form
$A_0\approx\ii\mu\sigma^3$, where $0\leq\mu\leq\pi/\beta$. Now make the
gauge transformation
$A_{\nu}' = g A_\nu g^{-1} - (\partial_\nu g)g^{-1}$, where $g=\exp(\ii\mu x^0 \sigma^3)$.
In the new gauge, $A_0'\rightarrow 0$ at infinity, and the gauge field is no
longer strictly-periodic, but satisfies
\[
  A_\nu'(x^0+\beta, x^1,x^2,x^3) = h A_\nu(x^0,x^1,x^2,x^3) h^{-1},
\]
where $h=\exp(i\mu\beta\sigma^3)$. This gauge choice is known as the algebraic
gauge. We can obtain 1-skyrmion chain configurations by computing the
holonomies of such
calorons in the algebraic gauge. Analytic expressions are known for all calorons
with unit instanton charge and vanishing monopole charge \cite{KB98}; so it is
feasible to study Skyrme chains using this family. Since we are mainly
interested in maximally-attractive chains, we restrict attention to the calorons
which yield $R^2=1$ but $R\neq1$; these are the calorons with $\mu=\pi/2\beta$.
By contrast, calorons with $\mu=0$, such as the Harrington-Shepard calorons
\cite{HS78}, give rise to aligned chains.

The calorons we are interested in are symmetric under rotations about an axis in
$\mathbb{R}^3$.  If we choose to evaluate holonomies along lines parallel to this
axis, the Skyrme chains will also have an $SO(2)$ symmetry: these will be
maximally-repulsive chains.  If, on the other hand, we evaluate holonomies
along lines perpendicular to the symmetry axis, we obtain maximally-attractive
chains.

We have implemented the Atiyah-Manton construction for maximally-attractive
chains
numerically, and evaluated the energies of the resulting skyrmion chains.
The family of calorons we used is parametrized by a scale parameter $\rho$ and a
period $\beta$.  The caloron with scale $\rho$ and period $\beta$ is in fact a
rescaling of the caloron with scale $\rho/\beta$ and period 1, and since the
components of the Skyrme energy behave simply under rescalings, it was sufficient
only to consider calorons with fixed $\beta=1$ and a range of values of $\rho$.

The holonomies were evaluated using the Runge-Kutta method.  We evaluated
energies
in a finite box $-L\leq x,y\leq L$, and extrapolated in both the box size and
the lattice spacing to obtain energies accurate to within $0.1\%$.  We also
calculated $B$ to check the accuracy of our method.  Our results are summarized
in Table~1.
\begin{table}
\begin{center}
\begin{tabular}{|l|l|l|l|}
\hline
$\rho$ & $12\pi^2E_2$ & $12\pi^2E_4$ & $B-1$ \\
\hline
0.2 & 10.56 & 509.2 & $2.0\times 10^{-4}$ \\
0.3 & 16.61 & 314.3 & $1.6\times 10^{-4}$ \\
0.4 & 23.20 & 213.7 & $1.1\times 10^{-4}$ \\
0.5 & 29.68 & 160.4 & $7.7\times 10^{-5}$ \\
0.6 & 35.40 & 135.4 & $6.3\times 10^{-5}$ \\
0.7 & 40.21 & 124.6 & $9.3\times 10^{-5}$ \\
0.8 & 44.25 & 119.6 & $1.9\times 10^{-4}$ \\
\hline
\end{tabular}
\caption{Caloron approximation}\label{Tab1}
\end{center}
\end{table}

In figure 1(a), we have plotted the minimum energy of this approximate
skyrmion chain, as a function of the period $\beta$. The graph was obtained by
interpolating the data in Table~1 to obtain $E_2$ and $E_4$ as polynomial
functions of $\rho$, and then minimizing the energy $E=E_2\beta + E_4/\beta$
with respect to variation in $\rho$. In particular, we see that the energy
of these caloron-derived configurations has its lowest value $E\approx1.16$
for period $\beta\approx2.1$.
\begin{figure}[htb]
\begin{center}
\includegraphics[scale=0.5]{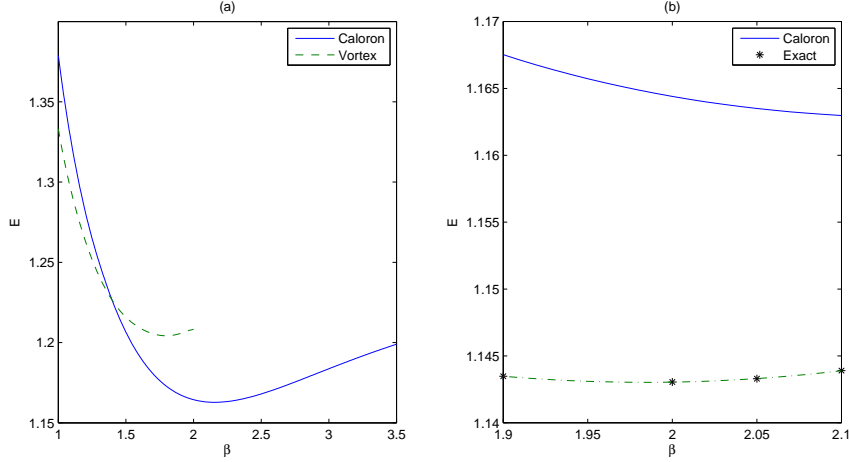}
\caption{The energy of a 1-skyrmion chain, versus the period $\beta$,
  for the caloron construction (solid curve), the vortex ansatz (dashed curve),
  and from a full 3D calculation (dot-dash curve). \label{fig1}}
\end{center}
\end{figure}
%


\section{The vortex ansatz}

In this section we describe an alternative ansatz for chains, which is based on
the idea that a chain, especially for small period, splits into constituents.
We define a Skyrme vortex to be a field of the form
\begin{equation}\label{vortex}
  U_v = \exp\left[\half(\theta - \nu z) \ii\sigma^3 \right]
    \exp\left[f(r)\ii\sigma^1\right]
    \exp\left[\half(\theta + \nu z) \ii\sigma^3 \right],
\end{equation}
where $(r,\theta)$ are polar coordinates in $\mathbb{R}^2$, and $\nu$ is a
positive constant. The profile function $f(r)$ is required to satisfy the
boundary conditions
$f(0)=\pi/2$ and $f(r)\rightarrow0$ as $r\rightarrow\infty$.  Note that
$U_v$ is smooth on $\mathbb{R}^3$, and periodic in $z$ with period $2\pi/\nu$.
The energy density of $U_v$ is
\begin{equation}
  \mathcal{E} = (f')^2 + \frac{\cos^2f}{r^2} + \nu^2\sin^2f
    +(f')^2\frac{\cos^2f}{r^2}+(f')^2\nu^2\sin^2f+\frac{\nu^2\sin^2f\cos^2f}{r^2}.
\end{equation}
The profile function $f(r)$ is chosen to solve the Euler-Lagrange equation
for $\int_0^\infty \mathcal{E} r\,dr$. But note, for example from the second
term in ${\cal E}$, that $U_v$ has infinite energy per unit period. This is
a consequence of the fact that $U_v$ has the non-constant form
\begin{equation}\label{vortex_infty}
  U_v \approx \exp(\ii\theta\sigma^3)
\end{equation}
as $r\to\infty$.
The Skyrme vortex (\ref{vortex}) has been used before, in a different context
\cite{KS87}: constructing vortex loops by taking finite lengths
of Skyrme vortex and joining their ends together.

Notice that the field $U(x,y,z) = U_v(x,-y,-z)$ winds in the opposite direction
to $U_v$ at infinity. So we can obtain a field which is constant at infinity
by taking a superposition of two vortices via the product ansatz
\[
    U = U_1 U_2
\]
with $U_1(x,y,z) = U_v(x-a,y,z)$ and $U_2(x,y,z) = U_v(x+a,-y,-z)$, where
$a$ is a positive constant. We can obtain a similar, but more symmetric,
field by using the relativized product ansatz \cite{NR88}
\[
  U = (U_1 U_2 + U_2 U_1)/\sqrt{({\rm det}(U_1U_2 + U_2 U_1)}.
\]
The superposition satisfies the boundary conditions of a maximally-attractive
chain with period $\beta=\pi/\nu$ and $R={\rm diag}(-1,-1,1)\in {\rm SO(3)}$,
and has the same symmetries.  The field resembles a pair of parallel
vortices separated by a distance $2a$, and its topological charge is 1.

When the separation $2a$ of the vortices is large, they attract each other, as
the following heuristic argument shows. Let $C>0$ be sufficiently large for the
approximation (\ref{vortex_infty}) to be valid for $r>C$, and let $a$ be
larger than $C$.  We can evaluate the energy of the superposition by
splitting $\mathbb{R}^2$
into three regions: the two discs of radius $C$ centred on the vortex locations,
and the exterior.  The energy within each disc tends to a constant as $a$ tends
to infinity.  The energy in the exterior diverges as $a\rightarrow\infty$;
a calculation shows that the leading contribution at large $a$ is
$4\pi\beta\ln a$.
So one can reduce the energy by reducing the separation $2a$, as claimed.

So far we have not justified our choice of superposition procedure: it is
important
to ask whether there is another way to superpose two vortices to obtain a lower
energy.  Again, we have a heuristic argument why our superposition is the right
thing to do, at least for large separation. Let $F$ denote the exterior of the
two discs $D_1$, $D_2$ of radius $C$ and centres $(x,y)=(\pm a,0)$, as before.
Let $\psi:F\rightarrow U(1)$ be a map such that $\psi|_{\partial D_1}$ has winding
number $1$ and $\psi|_{\partial D_2}$ has winding number $-1$.  We want to
minimize the energy
\[
   e = \beta \int_F \|\psi^{-1} d\psi \|^2 d^2x.
\]
The ansatz used above corresponds to taking
$\psi = \exp[\ii(\theta_1-\theta_2)]$,
where $\theta_1(x,y)$ is the angle between $(x-a,y)$ and the $x$-axis, and
$\theta_2(x,y)$ is the angle between $(x+a,y)$ and the $x$-axis. If the energy of
this field is close to the true minimum, then we know our ansatz is a good one.
Notice that the Skyrme term has disappeared from our energy functional; this is
because the Skyrme term evaluates to zero for any $U(1)$ field.
The easiest way to find the minimum energy is to stereographically project from
$\mathbb{R}^2$ to $S^2$; the energy $e$ is conformally-invariant, so we are
allowed to do this.  The stereographic projection can be chosen so that the
two circles are described by $\theta = \alpha$ and $\theta=\pi-\alpha$ in
spherical coordinates
$\theta\in[0,\pi]$, $\phi\in[0,2\pi)$, where $\sin(\alpha)=C/a$.  The energy
functional is now written as
\[
  e = \beta\int_0^{2\pi} \int_{\alpha}^{\pi-\alpha}\left( (\psi^{-1}\partial_\theta \psi)^2
   +\sin^{-2}\theta(\psi^{-1}\partial_\phi\psi)^2\right)\sin\theta\,d\theta\,d\phi.
\]
A Bogomolny argument shows that this energy is minimized by
$\psi(\theta,\phi)=\exp(i\phi)$; and the minimum energy is
$4\pi\beta\ln \cot(\alpha/2)$.
For large $a$, this agrees with our superposition, to leading order.

We have evaluated the energy of the superposition of two vortices for a range
of values of $\beta$ and $a$.  The energies were evaluated in a finite box, and
we extrapolated in the box size and the lattice spacing to obtain results
accurate
to within $0.1\%$.  We tried using both the product ansatz and the relativized
product ansatz, and found that the energies obtained agreed.  We also evaluated
the topological charge $B$ as a check on our methods.  Table~2 shows
the minimum energy of the superposition, together with the value of $a$ for
which this energy is attained. The energy is plotted, as a function of the
period $\beta$, in Figure~1(a).
\begin{table}
\begin{center}
\begin{tabular}{|l|l|l|l|}
\hline
$\beta$ & $12\pi^2E$ & $a$ & $B-1$ \\
\hline
1.0 & 157.9 & 1.13 & $5.4\times10^{-4}$ \\
1.1 & 153.1 & 1.15 & $3.4\times10^{-4}$ \\
1.2 & 149.6 & 1.16 & $3.1\times10^{-4}$ \\
1.3 & 147.1 & 1.18 & $2.1\times10^{-4}$ \\
1.4 & 145.2 & 1.20 & $2.0\times10^{-4}$ \\
1.5 & 144.0 & 1.21 & $1.5\times10^{-4}$ \\
1.6 & 143.2 & 1.22 & $1.5\times10^{-4}$ \\
1.7 & 142.8 & 1.23 & $1.2\times10^{-4}$ \\
1.8 & 142.6 & 1.24 & $1.1\times10^{-4}$ \\
1.9 & 142.8 & 1.24 & $9\times10^{-5}$ \\
2.0 & 143.1 & 1.25 & $9\times10^{-5}$ \\
\hline
\end{tabular}
\caption{Vortex approximation}\label{Tab2}
\end{center}
\end{table}


\section{Numerical Minimization.}

In order to assess the approximation schemes of the previous two sections,
and also to investigate the stability of the maximally-attracting 1-skyrmion
chain, we implemented a full 3-dimensional numerical minimization of the
energy $E$. The method involved a first-order finite-difference scheme for
$E$, with the spatial points $(x,y,z)$ being represented by a rectangular
lattice having lattice spacing $h$. The boundary condition $U=1$ was imposed
at $|x|=L$, $|y|=L$. The energy was minimized using a conjugate-gradient
method.  The errors in $E$ as a result of the finite
lattice spacing $h$ and finite size $L$ go like $h^2$ and $1/L^2$
respectively, and we obtained the $h\to0$, $L\to\infty$ limits by
extrapolation. The resulting values for $E$ have an error less than
$0.05\%$.

The first simulation looked at the maximally-attractive 1-skyrmion
chain, for each of the periods $\beta=1.9, 2.0, 2.05, 2.1$
(since all the indications are that the preferred period $\beta_{{\rm min}}$
is close to 2). The results are plotted in Figure 1(b), together with a
parabola fitted to the resulting four points; we see that the preferred
period is $\beta_{{\rm min}}=1.98$, and that the minimal energy is
$E_{{\rm min}}=1.143$.

For large period, we would expect the maximally-attractive 1-skyrmion chain
to be unstable to clumping, for the same reason that a finite collection of
separated 1-skyrmions will clump together. For small period, however,
the 1-skyrmion chain might be stable.  To investigate this, we first
looked at the periodic 2-skyrmion chain, where the initial configuration
was taken to be (a deformation of) a pair of skyrmions on the $z$-axis,
in the attractive channel. As before, the numerical simulation involved flowing
down the energy gradient, to reach a local minimum. The results are illustrated
in Figure~2, for two values of the period, namely $\beta=4$ and $\beta=5$.
\begin{figure}[htb]
\begin{center}
\includegraphics[scale=0.8]{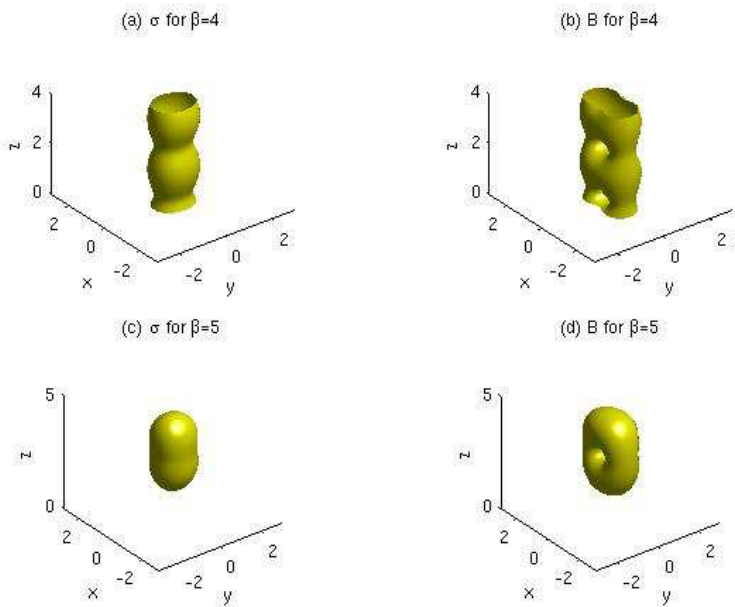}
\caption{The field component $\sigma$ and the charge density ${\cal B}$ for
   the 2-skyrmion chains with periods $\beta=4$ and $\beta=5$. \label{fig2}}
\end{center}
\end{figure}
The 1-skyrmion chain with period $\beta=2$ (essentially the
preferred period) corresponds to a
2-skyrmion chain with period $\beta=4$, and this 2-skyrmion chain turns out
to be stable. The minimal configuration is depicted in the upper row of
Figure~2. Subfigure (a) plots the function $\sigma=\half\tr(U)$, or rather
the surface $\sigma({\bf x})=0$;
subfigure (b) plots the charge density ${\cal B}$, or rather
the surface ${\cal B}({\bf x})=0.2\times\max{\cal B}$.

For the larger period
$\beta=5$, however, the two skyrmions coalesce to form a single
2-skyrmion, which (as one would expect from the $\mathbb{R}^3$ case) has a
toroidal shape. This is depicted in the lower row of Figure~2, where the same
quantities are plotted as in the upper row. The instability which leads
to this coalescence is rather weak: in fact, the energy of the toroidal chain
with period $\beta=5$ is only slightly ($0.1\%$) less than twice the energy
of a 1-skyrmion chain with period $2.5$.

Although the 1-skyrmion chain with $\beta=2$ is stable when considered
over two periods, one might expect that it will in turn be unstable to
clumping if examined over more than two periods. For example, there is evidence
suggesting that a 4-skyrmion chain (chain of $\alpha$-particles) might be
particularly favourable:
finite-length chains of $\alpha$-particles occur in $\mathbb{R}^{3}$,
at least when one modifies the system by adding a significant pion mass
\cite{BS05, BS06, BMS07}. We again used numerical minimization, this time
looking for the minimal-energy chain of 4-skyrmions, with various periods.
We use the periodicity condition which amounts to rotating neighbouring
$\alpha$-particles by $\pi$ about the periodic axis \cite{BMS07};
this is of the form (\ref{orientation}) with $R^2=1$ and $R\neq1$.
The $\alpha$-particle chain with period $\beta=8$ has has significantly
($4\%$) lower energy than four times the energy of a 1-skyrmion with
period $2$; the plot of its charge density in Figure~3(a) shows that, for
this value of the period, the $\alpha$-particles are quite well localized.
One can lower its energy by reducing the period, and so allowing
the $\alpha$-particles to move closer together: in Figure~3(b), we see the
$\alpha$-particle chain with period $\beta=3$, and its energy is
$6\%$ lower than the minimal value ($E=1.143$) for a 1-skyrmion chain.
\begin{figure}[htb]
\begin{center}
\includegraphics[scale=0.6]{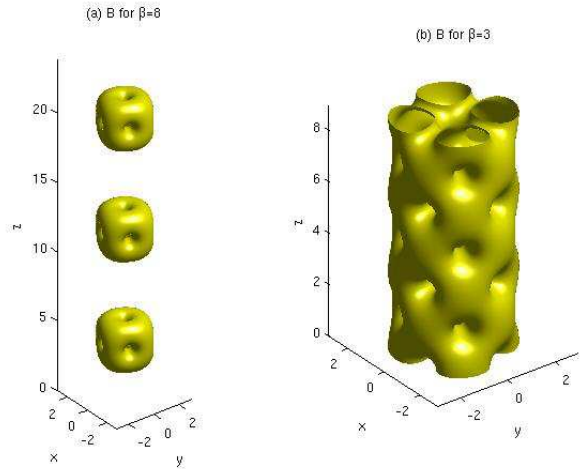}
\caption{The charge density ${\cal B}$ for 4-skyrmion chains with periods
    $\beta=8$ and $\beta=3$, over three periods. \label{fig3}}
\end{center}
\end{figure}
%


\section{Concluding Remarks}

From Figure 1 we see that the Atiyah-Manton construction gives a reasonably good
approximation to the minimal-energy 1-skyrmion chain, with energy only
$1\%$ above
its true value. The vortex-antivortex approximation is not quite as good, but
does emphasize the constituent structure of skyrmion chains. It appears to be
a common feature of topological soliton systems that soliton chains exhibit a
constituent structure when the soliton size is comparable to the period: the
solitons
fragment into fractional-charge objects, and this fragmentation occurs in a
transverse direction, so that rotational symmetry about the chain axis is lost.
These features are apparent from Figure 2(b), where one sees evidence of the
parallel vortex-antivortex pair. For low-period chains (more precisely, where
the period is small compared to the natural soliton size), one expects the
chain to resemble a vortex pair of this type.

The Skyrme crystal \cite{KS88, KS89, CJJVJ89} has an energy-per-baryon of
$E=1.036$, and one could construct
a skyrmion chain by cutting a chain out of this crystal, in other words by
truncating it in the $x$- and $y$-directions. A very thick chain obtained in this
way should have rather low energy (becoming lower as the chain became thicker).
Among thin chains (with transverse size comparable to the natural soliton size),
the chain of $\alpha$-particles may well be the lowest-energy solution. The
preliminary calculation reported in the previous section showed that its energy
is $E\approx1.07$ for period $\beta=3$. It would be worth making a more
comprehensive study of $N$-skyrmion chains for various $N$, for a wide range
of periods $\beta$,
and with various periodicity conditions; but that will require more intensive
computational effort than we have used in deriving the results reported here.


\bigskip\noindent{\bf Acknowledgments.}
This work was supported by the research grant ``Strongly Coupled Phenomena'',
and a research studentship, from the UK Science and Technology Facilities
Council.  DH is grateful to Dirk Sh\"utz for discussions.


\section{Appendix: Topology of Chains}

In this Appendix, we show that the topological charge $B$ is an integer,
even when the field is not strictly-periodic.  It is a special case of the 
following generalisation of the degree theorem, which appears to be new.

\medskip\noindent{\bf Theorem.}
Let $\Sigma$ be an $n$-dimensional compact manifold without boundary, with
volume form $\omega$, and with $H^{n-1}(\Sigma)=0$ and $H^n(\Sigma)=\mathbb{Z}$.
Let $M$ be an $(n-1)$-dimensional manifold such that
$H^n(M\times S^1)=\mathbb{Z}$.
Suppose that $SO(2)$ acts on $\Sigma$ and that $\omega$ is $SO(2)$-invariant.
Fix an element $\sigma\in SO(2)$ and let
$\phi:M\times \mathbb{R} \rightarrow \Sigma$ be a map satisfying
\[
  \phi(x,y+\beta) = \sigma\phi(x,y) \; \forall x\in M, y\in\mathbb{R}.
\]
Then $\phi$ has an integer degree, computed by the integral
\[
  deg(\phi) = \frac{1}{Vol(\Sigma)} \int_0^\beta \int_{M} \phi^* \omega.
\]
Furthermore, $deg(\phi)$ is independent of the choice of $SO(2)$-invariant
volume form $\omega$.

\medskip\noindent{\bf Proof.}
The idea of the proof is simple: we deform $\phi$ to a strictly-periodic map
using the $SO(2)$ action, and show that the integral is unchanged by this
deformation.  First, we introduce some notation: we write the $SO(2)$ action
as $R_s:\Sigma\rightarrow\Sigma$, with $s\in \mathbb{R}/\mathbb{Z} \cong SO(2)$.
Let $X\in T\Sigma$ be the associated vector field.  Let
$t_0\in\mathbb{R}/\mathbb{Z}$ be such that $R_{t_0}=\sigma^{-1}$, and let
$t(x,y)=t_0y/\beta$ be a function on $M\times\mathbb{R}$.  We define a
deformation
\[
  \tilde{\phi}(x,y) = R_{t(x,y)}(\phi(x,y)) ,\, x\in M,y\in\mathbb{R}.
\]
Then $\tilde{\phi}$ is a strictly periodic map, hence has a degree computed by
\[
  deg(\tilde{\phi}) =
    \frac{1}{Vol(\Sigma)} \int_0^\beta \int_{M} \tilde{\phi}^* \omega.
\]
Now we show that
\[
  \int_0^\beta \int_{M} \tilde{\phi}^* \omega = \int_0^\beta \int_{M} \phi^* \omega.
\]
For any form $\theta\in\Lambda^*\Sigma$, one can show that
\[
  \tilde{\phi}^*\theta = \phi^*R_t^*\theta + \phi^*(i_X R_t^*\theta)\wedge dt.
\]
Here $i_X$ denotes the inner derivative of a form.  In the particular case
$\theta=\omega$, one has $R_t^*\omega=\omega$ (because the volume form is
$SO(2)$-invariant).  Hence
\[
  \tilde{\phi}^*\omega = \phi^*\omega + \phi^*(i_X \omega)\wedge dt.
\]
By Cartan's formula, we have
\[
   L_X \omega = i_Xd\omega + di_X\omega,
\]
where $L_X$ denotes the Lie derivative.  We see immediately that $d\omega=0$,
because $\omega\in\Lambda^n\Sigma$.  On the other hand, $L_X\omega$ must
vanish since $\omega$ is $SO(2)$-invariant.  It follows that $i_X\omega$
is closed.  Since $H^{n-1}(\Sigma)=0$, $i_X\omega$ is exact, in other words,
there exists a $\mu\in\Lambda^{n-2}\Sigma$ such that $i_X\omega=d\mu$.
Therefore
\begin{eqnarray*}
\tilde{\phi}^*\omega &=& \phi^*\omega + \phi^*(d\mu)\wedge dt \\
    &=& \phi^*\omega + d(\phi^*\mu\wedge dt).
\end{eqnarray*}
Integrating and applying Stoke's theorem, we obtain
\[
  \int_0^\beta \int_{M} \tilde{\phi}^* \omega = \int_0^\beta \int_{M} \phi^* \omega +
     \left[ \int_{M} \phi^*\mu\wedge dt \right]_0^\beta.
\]
The boundary term vanishes since $t$ is constant on the domain of
integration, so we have the desired result.

That $deg(\phi)$ is independent of the choice of volume form follows from
the corresponding property of the classical degree.  If $\omega'$ is any
other $SO(2)$-invariant volume form, then
\begin{eqnarray*}
\int_0^\beta \int_{M} \phi^* \omega' &=& \int_0^\beta \int_{M} \tilde{\phi}^* \omega'\\
   &=& \int_0^\beta \int_{M} \tilde{\phi}^* \omega \\
   &=& \int_0^\beta \int_{M} \phi^* \omega.
\end{eqnarray*}


\end{document}